\documentstyle{mnrastmp} 

\begin{document} 
\Received{2014}  \Accepted{2014} 

\title{The North Polar Spur and Aquila Rift}  
\author{Yoshiaki {\sc SOFUE}}
\affil{Institute of Astronomy, University of Tokyo, Mitaka, Tokyo 181-0015,
 sofue@ioa.s.u-tokyo.ac.jp} 

\KeyWords{ISM: individual objects (North Polar Spur, Aquila Rift) -- ISM: molecules -- ISM: kinematics and dynamics -- Galaxy: center -- local interstellar matter -- X-rays: diffuse background -- radio lines: ISM}

\maketitle

 \begin{abstract}
Soft X-ray intensity at 0.89 keV along the North Polar Spur is shown to follow the extinction law due to the interstellar gas in the Aquila Rift by analyzing the ROSAT archival data, which proves that the NPS is located behind the rift. The Aquila-Serpens molecular clouds, where the X-ray optical depth exceeds unity, are shown to have a mean LSR velocity of $ v=7.33 \pm 1.94 {\rm km~s}^{-1}$, corresponding to a kinematic distance of $ r=0.642 \pm 0.174 ~{\rm kpc}$. Assuming a shell structure, a lower limit of the distance to NPS is derived to be $1.01\pm 0.25$ kpc, with the shell center being located farther than 1.1 kpc. Based on the distance estimation, we argue that the NPS is a galactic halo object.
\end{abstract}

\def\kms{km s$^{-1}$} 
\def\Msun{M_{\odot \hskip-5.2pt \bullet}} 
\def\msun{$M_{\odot \hskip-5.2pt \bullet}$} 
\def\deg{^\circ}
\def\htwo{H$_2$} \def\vr{v_{\rm r}}
\def\be{\begin{equation}} \def\ee{\end{equation}} \def\bc{\begin{center}}
\def\ec{\end{center}} \def\bf{\begin{figure}} \def\ef{\end{figure}}  
\def\hcm2{{\rm H~cm}^{-2}}

\section{Introduction}

The North Polar Spur is defined as the prominent ridge of radio continuum emission, emerging from the galactic plane at $l\sim 20^\circ$ toward the north galactic pole (Haslam et al. 1982; Page et al. 2007). The spur is also prominent in soft X-rays as shown in the Wiskonsin and ROSAT all sky maps (McCammon et al. 1983; Snowden et al. 1997).

Discovery of the Fermi $\gamma$-ray bubbles (Su et al. 2010) has drawn attention to energetic activities in the Galactic Center, and possible relation of the NPS to the Fermi bubbles has been pointed out (Kataoka et al. 2013; Mou et al. 2014). The underlying idea is that the NPS is a shock front produced by an energetic event in the Galactic Center with released energy on the order of $\sim 10^{55-56}$ ergs  (Sofue 1977, 1984, 1994, 2000;  Bland-Hawthorn et al. 2005). In this model the distance to NPS is assumed to be $\sim 7$ kpc.

On the other hand, there have been traditional models to explain the NPS by local objects such as a nearby supernova remnant, hypernova remnant, or a wind front from massive stars (Berkhuijsen et al. 1971; Egger and Aschenbach 1995; Willingale et al. 2003;  Wolleben 2007;  Puspitarini et al. 2014). In these models the distance to NPS is assumed to be several hundred pc.
 
Hence, the distance is a crucial key to understand the origin of NPS. In this paper we revisit this classical problem, and give a constraint on the distance of NPS, following the method proposed by Sofue (1994). 
 
We adopt the common idea as the working hypothesis that the NPS is a part of a shock front shell composed of high-temperature plasma and compressed magnetic fields, emitting both the X-ray and synchrotron radio emissions. For the analysis, we use the archival data of ROSAT All Sky X-ray Survey (Snowden et al 1997), Argentine-Bonn-Leiden All Sky HI Survey (Kalberla et al. 2003), and Colombia Galactic Plane CO Survey (Dame et al. 2001).

\section{The North Polar Spur in Soft X-Rays}

Figure  1  shows a color-coded map of X-ray intensities in a $\pm 50^\circ$ region around the Galactic Center obtained by using the ROSAT archival FITS data, where red represents the R2 (0.21 keV) band intensity, green for R5 (0.89 keV), and blue for R7 (1.55 keV). The dashed lines trace the loci along which the analysis of X-ray absorption was obtained in this paper. The central locus below $b\sim 15^\circ$ was drawn along the radio continuum spur traced in the 1.4 GHz map by Sofue and Reich (1979). The white dashed line is a contour along which the optical depth of R5 band emission due to the interstellar gas is equal to unity.

\begin{table}
\caption{ROSAT X-ray bands and critical column density of H atoms (Snowden et al. 1997).} 
\begin{tabular}{lllll}  
\hline
$i$ &Band& $E$ range & Mean $E$ & $N_i~(\tau_i=1)$ \\
&& keV  & keV &  $10^{20}$ H cm$^{-2}$ \\
\hline
1&R1 &0.11  -- 0.28 &0.195 &0.73 \\
2&R2 &0.14  -- 0.28 &0.21 & 1.41\\ 
4&R4 &0.44 -- 1.01 &0.725 &23.0 \\
5&R5 &0.56 -- 1.21 &0.885 &27.\\
6&R6 &0.73 -- 1.56 &1.145 &36.\\
7&R7 &1.05 -- 2.04 &1.545 &90.\\
\hline
\end{tabular}
\end{table} 

Here, we chose R2, R5 and R7 bands in order to cover as wide range of X-ray energies as possible to see the variation of optical depths corresponding to strongly varying interstellar gas densities. We used R5 band as the representative of R4, R5 and R6 bands, because the three bands are largely overlapping. The X-ray emission from the NPS is shown to be originating from hot plasma of temperature $\sim 10^{6.3} - 10^{6.5}$ K (Snowden et al. 1997). Table 1 shows the energy bands of ROSAT observations.

\begin{figure}  
\bc
\includegraphics[width=7cm]{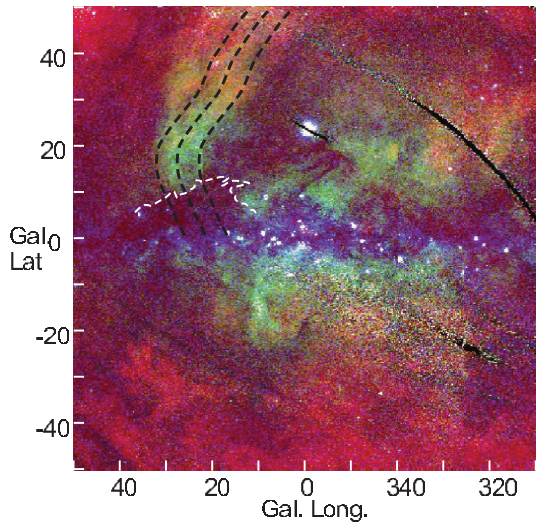}   
\ec
\caption{
Color coded X-ray map with red for R2 (0.21 keV), green for  R5 (0.89 keV), and blue for R7 (1.55 keV), as obtained using the ROSAT archival data (Snowden et al. 1997). The three dashed lines trace the North Polar Spur at the outer edge, peak intensity ridge, and inner region, along which the absorption analysis is obtained. The white dashed line indicates a contour along which the optical depth of R5 (0.89 keV) band emission due to the interstellar gas is equal to unity ($\tau_5=1$, or $N=N_5$).} 
\end{figure}

\begin{figure} 
\bc
\includegraphics[width=7cm]{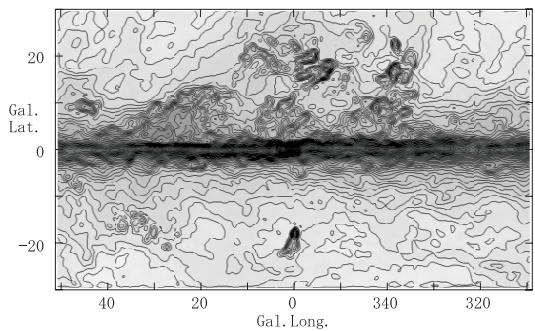}   
\ec
\caption{Hydrogen column density, $N = N({\rm HI}) + 2N({\rm H}_2)$, produced from the HI and CO line survey data by Kalberla et al (2003) and Dame et al. (2001).  Contour levels are logarithmic from 0 to 6686$\times 10^{23}$ H cm$^{-2}$ by exp(1/50) steps. }
\end{figure}

\section{The Aquila Rift composed of HI and \htwo~ Gases} 

The Aquila rift was originally known as a giant dark lane dividing the starlight Milky Way by heavy extinction (Weaver 1949; Dobashi et al. 2005: Arendt et al. 1998). We analyze the distribution of interstellar HI and \htwo ~ gases in the Aquila rift, and investigate the relation to absorption features of the X-ray NPS. We use the Leiden-Argentine-Bonn all-sky HI survey (Kalberla et al. 2005 ) and Colombia galactic plane CO survey (Dame et al. 2001). 

Figure 2 shows the distribution of column density of total hydrogen atoms $N=N({\rm H})+ 2N({\rm H_2})$ obtained from integrated intensities of the HI and CO lines. The integrations were obtained in the whole velocity ranges: HI line from $-200$ to +200 \kms, and CO line from $-300$ to $+300$ \kms. Hence, the map shows the column density along the entire line of sight. Prominent in these maps is the tilted ridge of HI and H$_2$ gases associated with the Aquila Rift. 

The hydrogen column density along the line of sight was calculated by
\be
N=C_{\rm HI} I_{\rm HI}+2 C_{\rm H2} I_{\rm CO},
\label{eqnint}
\ee
where I$_{\rm HI}$ and $I_{\rm CO}$ are the integrated intensities of HI and $^{12}$CO($1-0$) lines, and we adopt the conversion factors of $C_{\rm HI}=1.82\times 10^{18} [{\rm H~cm}^{-2}/{\rm K~km^{-1}}]$ and  $ C_{\rm H2}=2.0\times 10^{20} [{\rm H_2~ cm}^{-2}/ {\rm K~km^{-1}}]$. 
 
\section{Absorption of X-rays by Aquila Rift}
 
\subsection{Latitude variation of X-ray optical depth}  

Table 1 and figure 3 show energy dependency of the critical column density, $N_i$, at which the optical depth defined by 
\be
\tau_i=N/N_i,
\label{taui}
\ee
is equal to unity ($i=1,$ 2, ..., 7 corresponding to R1, R2, ..., R7, respectively), as obtained from Snowden et al. (2007). Typical critical densities used in this paper are $N_2=1.4\times 10^{20}{\rm H~cm}^{-2}$, $N_5=27.0 \times 10^{20}{\rm H~cm}^{-2}$, and $N_7=90.0\times 10^{20}{\rm H~cm}^{-2}$.  

\begin{figure} 
\bc
\includegraphics[width=7cm]{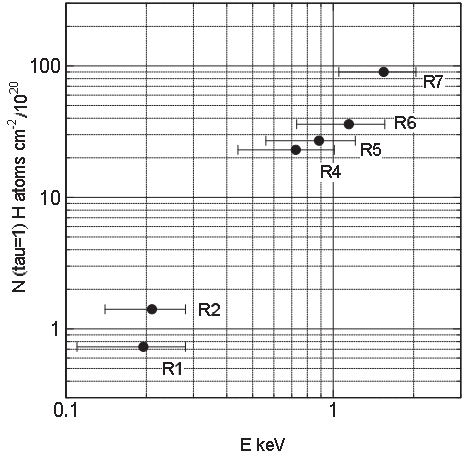}     
\ec
\caption{Critical hydrogen column density $N_i$ as a function of X-ray energy plotted for the ROSAT energy bands obtained by using the data by Snowden et al (1997).}
\end{figure} 

If X-rays in the $i$-th band are emitted at a distance $x$ with an intrinsic intensity $I_i^0$, they are absorbed by the intervening interstellar gas, and yield the observed intensity 
\be I_i=I_i^0 {\rm exp}(-\tau_i)+I_i^f, \ee 
where $I_i^f$ is the foreground emission. 

In figure 4 we plot R5-band intensities (ROSAT count rate per second per arc minute) against latitude in comparison with the total column density of hydrogen atoms along the three dashed loci indicated in figure  1 . The middle panel shows the plot along the NPS peak ridge (central locus), while the top panel for the eastern locus at $\delta l=+5^\circ$ along the outer edge of NPS, and the bottom for the western (inner) locus at $\delta l=-5^\circ$.  

Each plotted value is the mean of R5 band count rates in a box of extent $(\Delta l \times \Delta b)=(2^\circ \times 1^\circ)$ at every $1^\circ$ latitude interval along the loci. Long thin bars represent standard deviation $s_{\rm d}$ about the mean value in each box. Short thick bars are standard errors, $s_{\rm e}=s_{\rm d}/\sqrt{n}$, with $n\sim 55$ being the number of data points in each box.

The figure shows that the R5-band X-ray intensity is inversely correlated with the H column density. The intensity along the NPS peak ridge (middle panel) starts to decrease at $b=15^\circ$ showing a shoulder-like drop, and falls by $1/e$ at $b=11^\circ$, at which the H atom column density is $N=2.7 \times 10^{21} \hcm2$, coinciding with the threshold value $N_5$ corresponding to $\tau_5=1$. Similarly, a shoulder-like drop appears at $b\simeq 10^\circ$ along the outer edge (top panel), and at $b\simeq 16^\circ$ along the inner locus (bottom panel). Thus, the lower is the longitude, the higher is the latitude of the shoulder position. 

The steepest decreasing point of X rays in each panel coincides with the position at which the H column density is equal to the critical value $N=N_5=2.7\times 10^{21}$ H cm$^{-2}$, where $\tau_5=1$. Namely, R5-band X-rays of NPS fade away near the white-dashed contour in figure  1  in coincidence with the distribution of dark clouds in the Aquila Rift.

\begin{figure}   
\bc
\includegraphics[width=7cm]{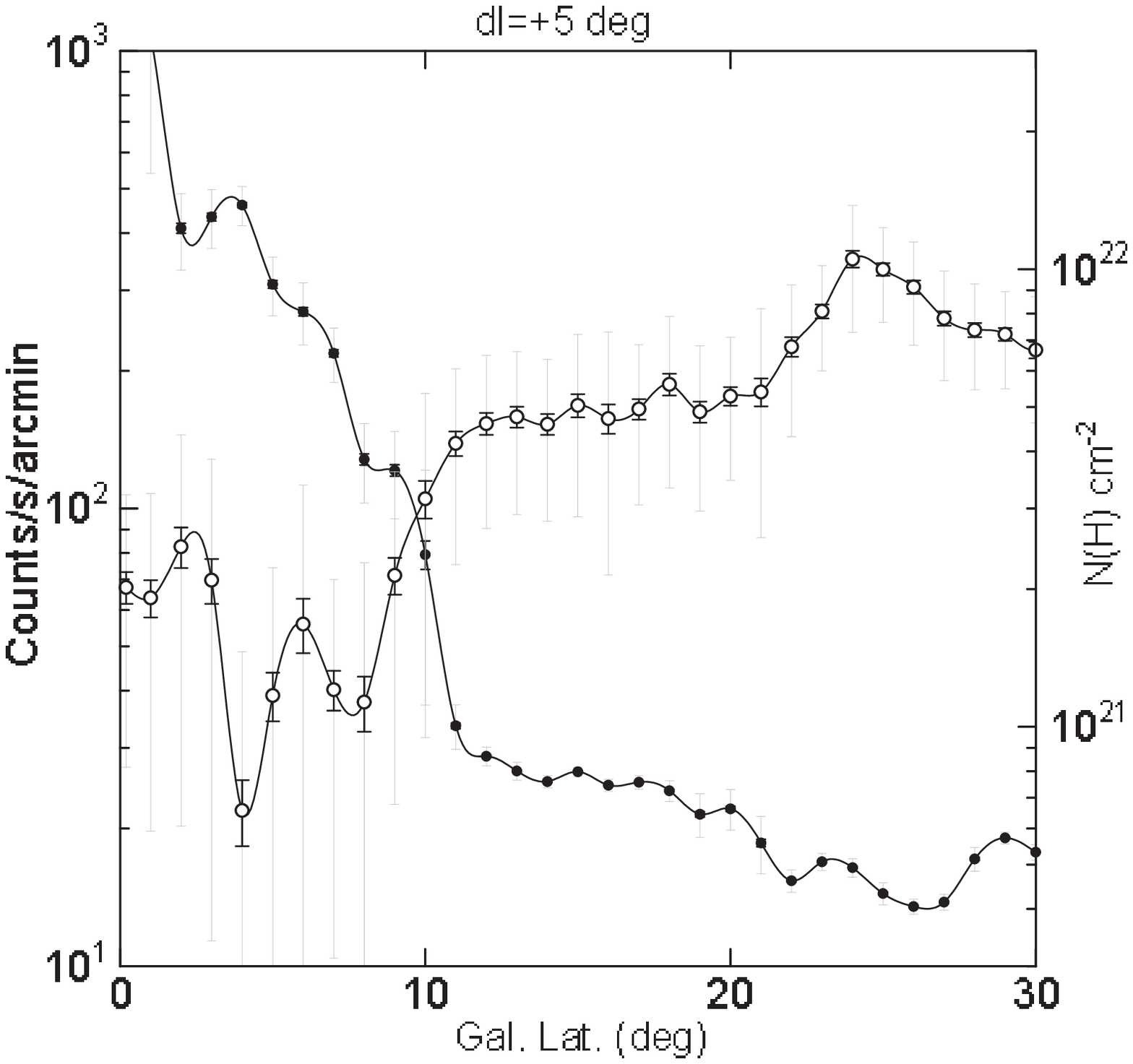}  
\includegraphics[width=7cm]{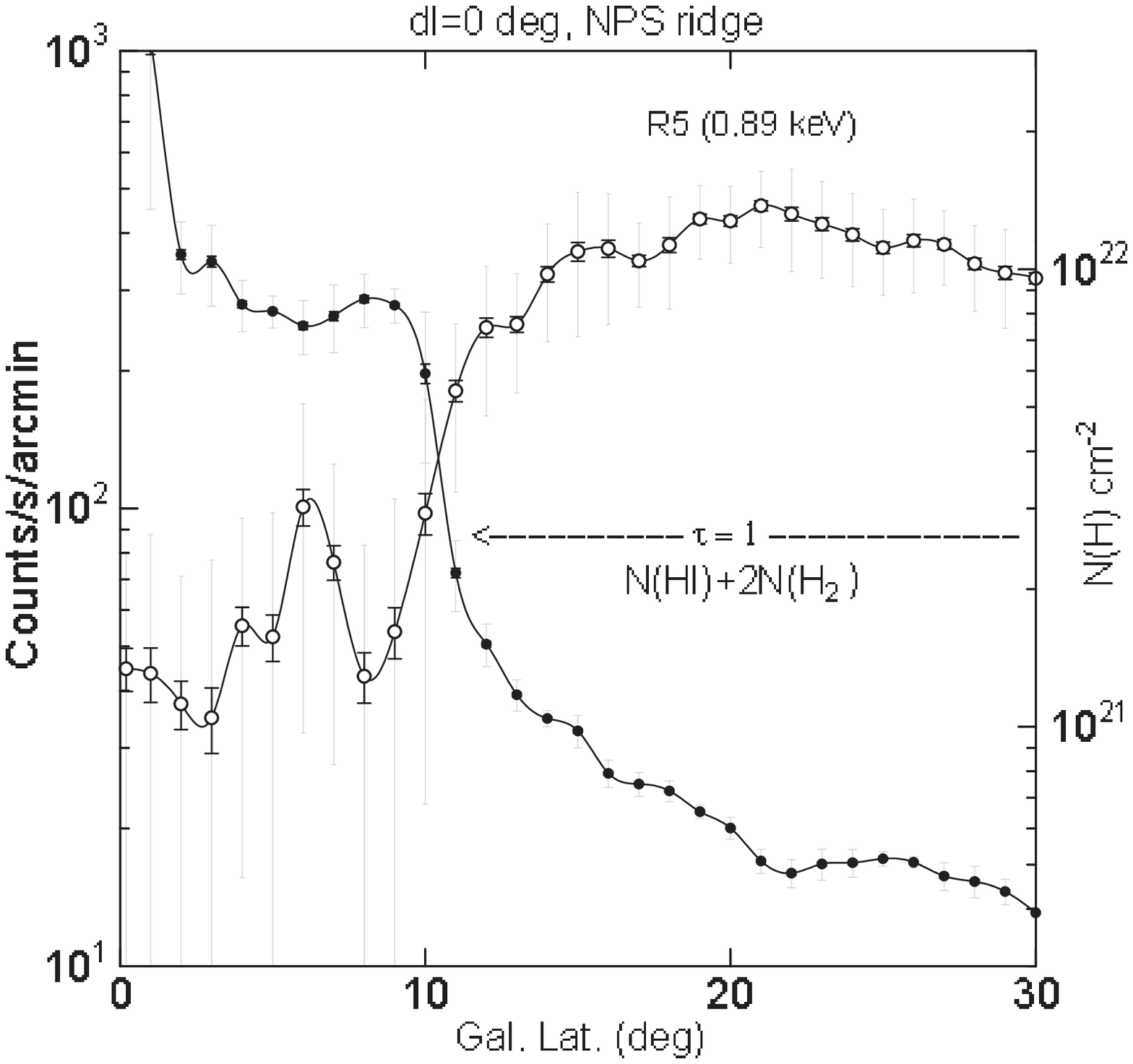}  
\includegraphics[width=7cm]{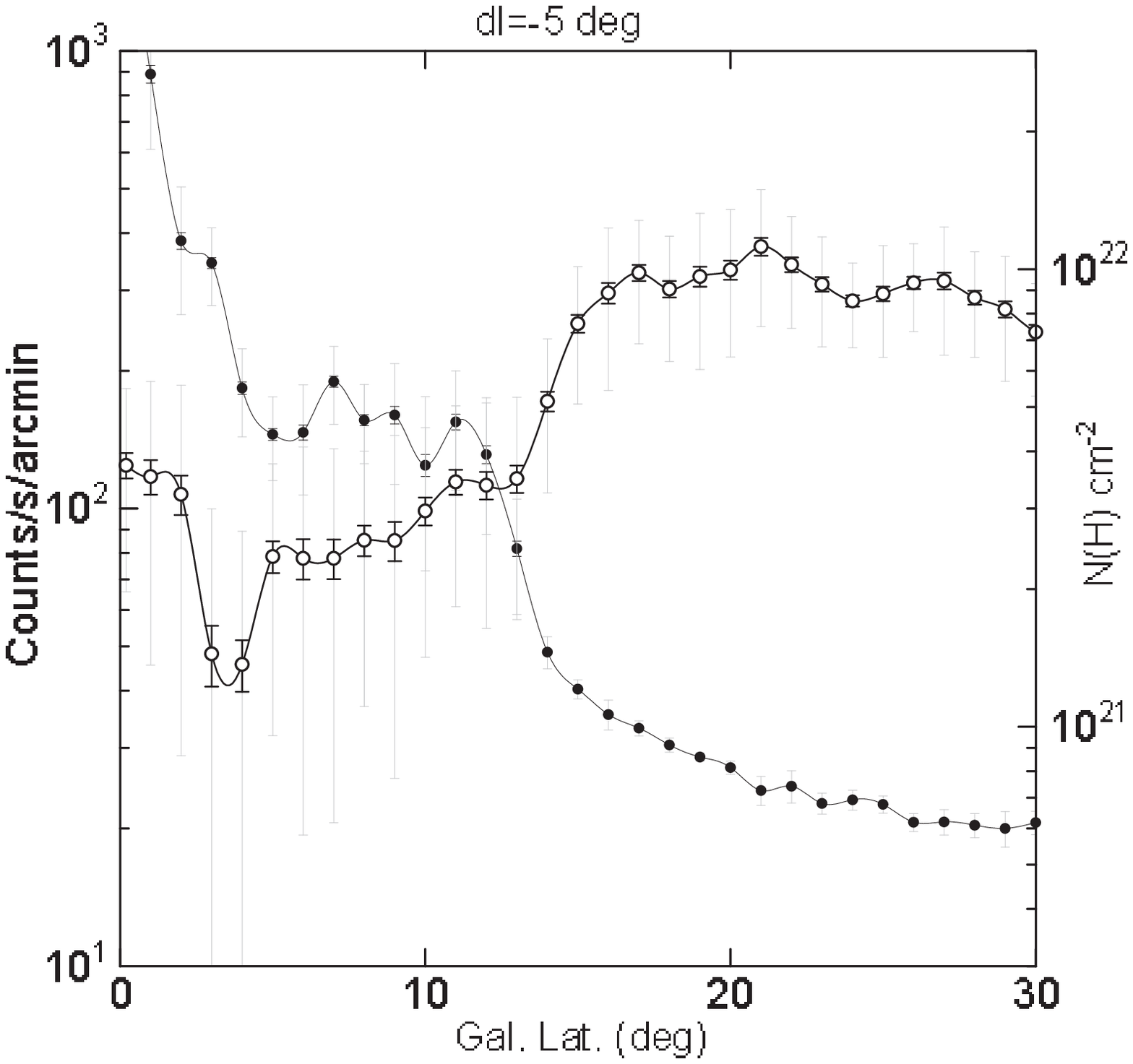}    
\ec
\caption{R5 band (0.89 keV) intensity (in ROSAT count rate s$^{-1}$ arcmin$^{-1}$) compared with total H column density, $N=N({\rm HI}) + 2 N({\rm H}_2)$, along the dashed lines in figure  1 ) plotted against $b$. From top to bottom: outer edge shifted by $\delta l=+5^\circ$ to the east, along the peak ridge at $\delta l=0^\circ$,  and inside NPS at $\delta l=-5^\circ$. Long and short bars represent $s_{\rm d}$ and $s_{\rm e}$, respectively.}
\end{figure} 

In figure 50mm we show the variation of R5 and R2 band intensity as a function of galactic latitude along the spur ridge. The X-ray intensity is normalized to the value at $b=45^\circ$ in each band. We also plot radio continuum intensity at 2.3 GHz using the Rhodes radio survey data by Jonas et al. (1998) after background filtering (Sofue and Reich 1979). The R2 intensity is strongly absorbed already at latitudes as high as $45^\circ$, where the H atom column density is $N\sim 4\times 10^{20} {\rm H~cm}^{-2}$ exceeding the threshold value of $N_2=1.4 \times 10^{20} \hcm2$, indicating $\tau_2>3$. 

\begin{figure} 
\bc
\includegraphics[width=7cm]{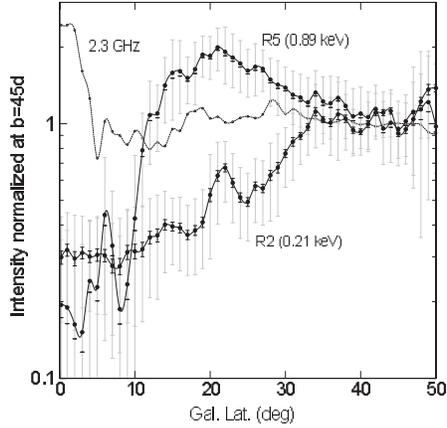}   
\ec
\vskip -5mm
\caption{Variations of relative X-ray intensities in R2 and R5 bands, and 2.3 GHz radio continuum brightness plotted along the NPS ridge against galactic latitude. Intensities are normalized at $b\simeq 45^\circ$. Long and short bars represent $s_{\rm d}$ and $s_{\rm e}$, respectively.}
\end{figure}

\subsection{Extinction law along the NPS}

Figure 6 shows R5 band intensity and H column density plotted against latitude in linear scale. We also show an extinction curve corresponding to the hydrogen column density, where the plotted values are proportional to exp$(-\tau_5)$ and normalized to the R5-band intensity at $b\sim 20^\circ$. Here, $\tau=N/N_5$ and $N_5=27.0 \times 10^{20}$ H cm$^{-2}$. The X-ray intensity follows almost exactly the extinction curve.

\begin{figure}   
\bc
\includegraphics[width=7cm]{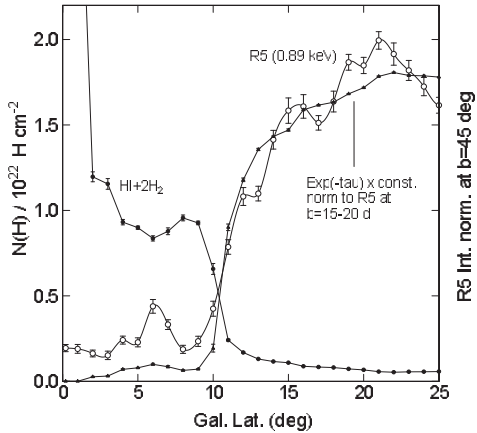}   
\ec
\caption{Relative R5 band intensity (circle) compared with the total hydrogen column density (dot) at $b<25^\circ$. Triangles show the extinction profile proportional to exp($-\tau_5$) normalized to the R5-band intensity at $b\sim 20-25^\circ$. Bars are standard errors.}
\end{figure} 

In figure 7, we plot R5 band intensity along NPS at $b<30^\circ$ as a function of the optical depth $\tau_5$. Here, the minimum value of intensity at $b\sim 5^\circ$ was subtracted as the foreground emission $I_5^{\rm f}$, and the intensity is normalized by the maximum value at $b\sim 20^\circ$. The inserted line shows the extinction law following an equation
\begin{equation}
I_5 =I_0 {\rm exp}(-\tau_5),
\end{equation}
where $I_0$ is the maximum intensity.

\begin{figure}   
\bc
\includegraphics[width=7cm]{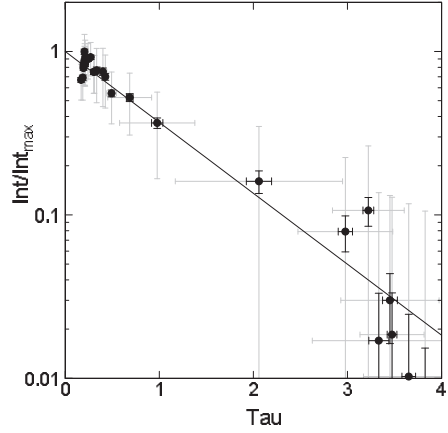} 
\ec
\caption{R5 band intensities at $b < 30^\circ$ plotted as a function of optical depth. The intensity is normalized by the maximum value at $b\sim 20^\circ$ with the minimum value at $b\sim 5^\circ$ being subtracted. The optical depth was calculated by $\tau=N/N_5$ with $N_5=27.0 \times 10^{20}$ H cm$^{-2}$. The straight line represents the extinction law, $I/I_0= {\rm exp}(-\tau_5)$. Long and short bars denote $s_{\rm d}$ and $s_{\rm e}$, respectively}
\end{figure}

The observed relations shown in figures 40mm, 50mm, 6 and 7 indicate that the X-ray intensity follows the simplest extinction law, and show that the NPS is indeed absorbed by the interstellar gas in the Aquila Rift region. The facts also prove that the NPS is an intrinsically continuous structure emitting X-rays even below latitudes at which the optical depth is greater than unity. 

This is consistent with our working hypothesis that the NPS is a continuous structure tracing the radio NPS. On the other hand, it denies the possibility that NPS has a real emission edge near $b\sim 11^\circ$, in which case the correlations in figures 6 and 7 are hard to be explained.

\section{Distance of Absorbing Neutral Gas}

\subsection{Kinematical distance from LSR velocity}  

A lower-limit distance of NPS can be obtained by analyzing the kinematical distance of the absorbing HI and molecular gases. The distance of the gas $r$ is related to the radial velocity  $\vr$ as   
\be
r=\vr/(A ~ {\rm sin}~2l ~ {\rm cos}~b),
\label{vr_r}
\ee
where $A$ is the Oort's constant. We here adopt the IAU recommended value $A=14.4$ \kms kpc$^{-1}$.

Figure 8cm shows CO line longitude-velocity (LV) diagrams at several galactic latitudes in the Aquila rift region. The diagrams show a prominent clump at around $\vr\sim 7$ \kms and $l\sim 25^\circ$. A corresponding feature is recognized in the HI LV diagram as a valley of emission at the same LV position at $b\simeq 5^\circ$, indicating that the CO Aquila rift is surrounded by an HI gas envelope. However, the HI diagrams were too complicated for distance estimation of the Aquila rift, and were not used here.

\begin{figure}  
\bc
\includegraphics[width=8cm]{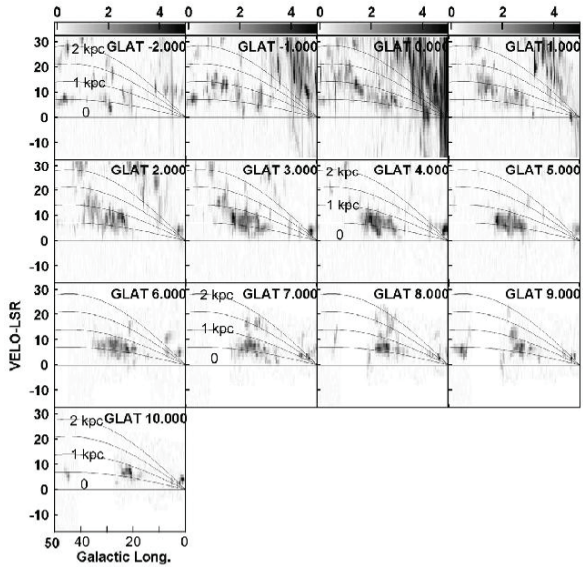}  
\ec
\caption{CO-line LV diagrams for $l=0^\circ$ to $50^\circ$ at different latitudes from data by Dame et al. (2001).  Inserted thin lines indicate kinematic LSR velocities corresponding to distances 0, 0.5, 1, 1.5 and 2 kpc from the Sun. This figure, therefore, represents approximate cross sections of the Aquila rift at different latitudes as seen from the galactic north pole. }
\end{figure}

In the CO LV diagram (figure 8cm), we insert thin lines showing velocities corresponding to distances of 0, 0.5, 1, 1.5 and 2 kpc, as calculated by equation \ref{vr_r}. The major CO Aquila rift is thus shown to be located at a distance of $r=0.6-0.7$ kpc with the nearest and farthest edges being at 0.3 and  1 kpc, respectively. At $b\sim 8^\circ$ an extended spur, reaching $\vr \sim 20$ \kms, seems to be associated with the major ridge, whose kinematical distance is estimated to be as large as $r \sim 1.8$ kpc. This figure, therefore, represents cross sections of the Aquila rift at different latitudes as seen from the galactic north pole.

Figure 9 shows CO line spectra toward the densest region in the Aquila Rift. Each line profile includes velocity information mixed with the cloud motion and velocity dispersion. The velocity dispersion is on the order of a few \kms, but is difficult to separate from the kinematical component.

\begin{figure} 
\includegraphics[width=2.5cm]{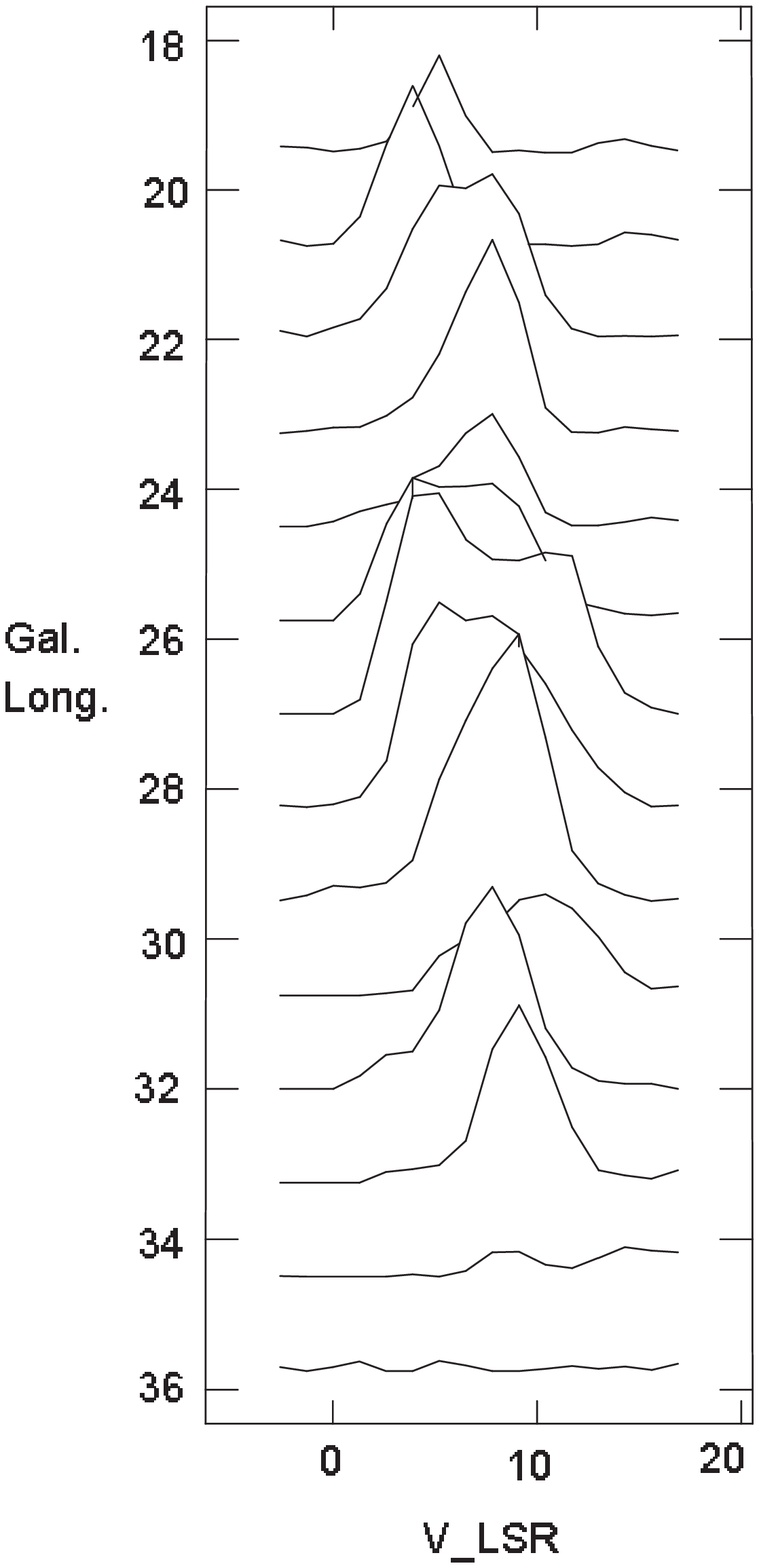}  
\includegraphics[width=1.8cm]{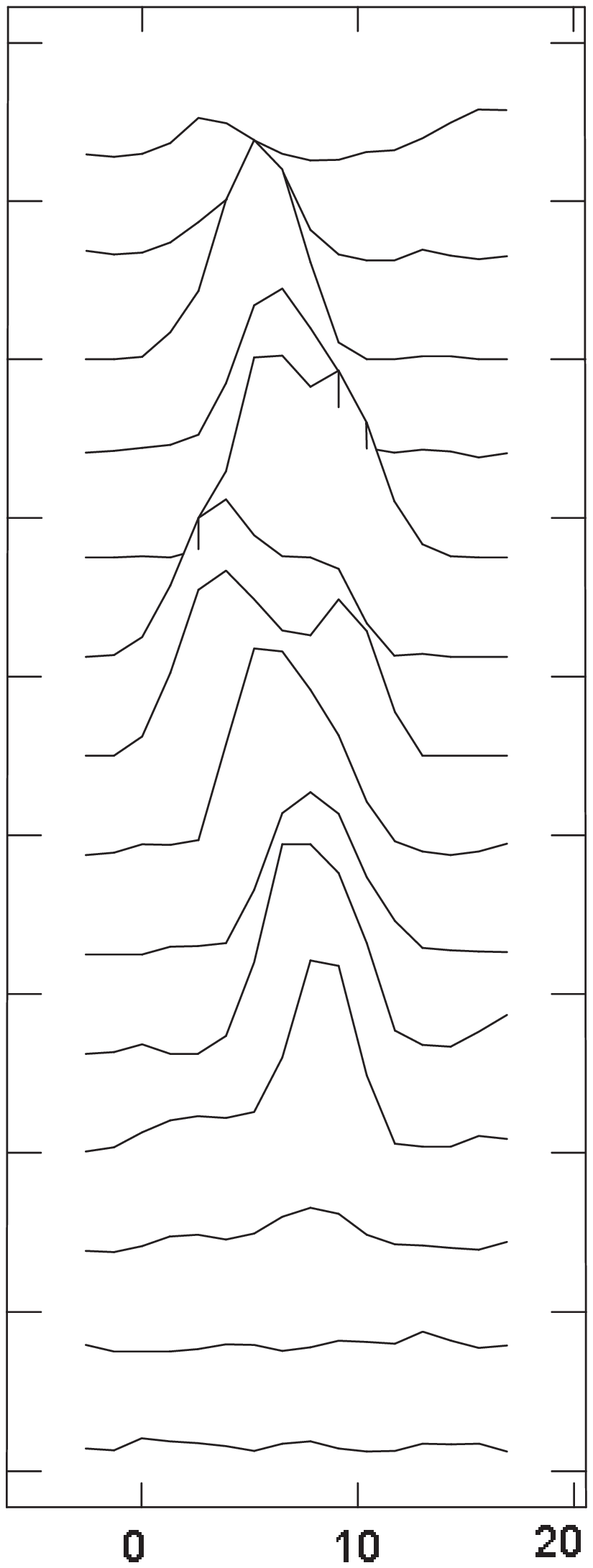}  
\includegraphics[width=1.8cm]{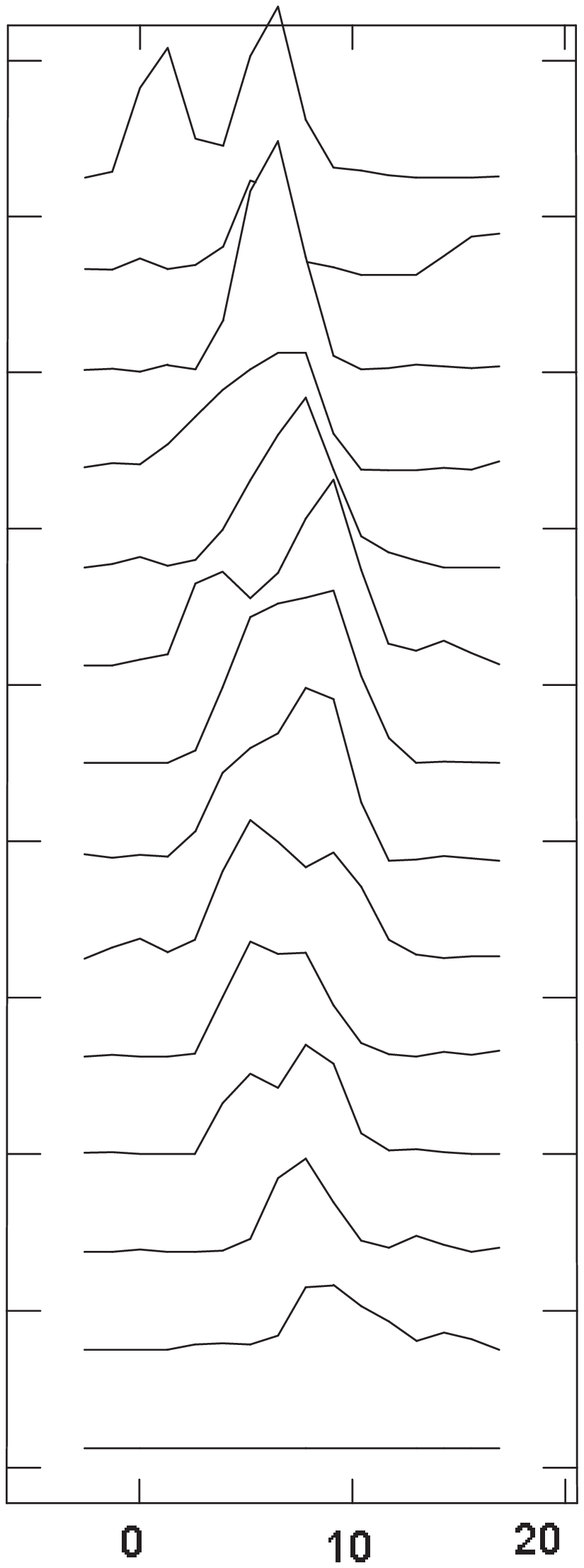}  
\includegraphics[width=1.8cm]{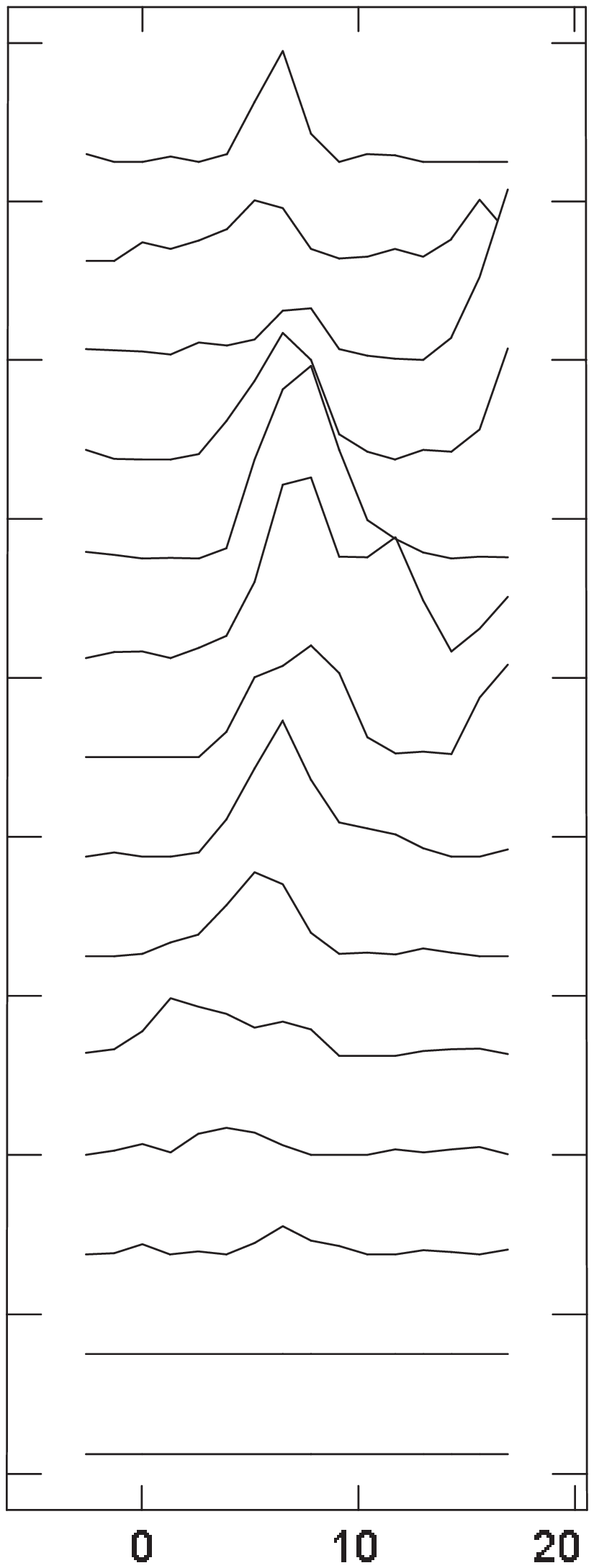}   
\caption{CO line spectra at latitudes $b=3^\circ.62, ~4^\circ.87, ~6^\circ.12$ and $7^\circ.37$ (from left to right) with longitude interval of about $2^\circ$. The ordinate tick interval corresponds to the CO-line brightness temperature of 2 K.}
\end{figure}

We, therefore, used peak velocities of individual spectra.  We read the peak velocities in the spectra shown in figure 9 from $l=22^\circ$ to $32^\circ$ and from $b=3^\circ.62$ to $7^\circ.36$. If a spectrum had two peaks, we adopted two corresponding velocities. Thus we obtained 44 peak velocities. We then calculated kinematical distance for each measured velocity using equation \ref{vr_r}. The measured velocities and calculated distances were averaged to obtain a mean velocity and distance as follows:
\be
v=7.33 \pm 1.94 (0.29) {\rm km~s}^{-1},
\ee
and 
\be
r=0.642 \pm 0.174 (0.026) ~{\rm kpc},
\ee 
where the former errors are the standard deviation $s_{\rm d}$, and the latter in the parenthesis are the standard errors of the mean values $s_{\rm e}=s_{\rm d}/\sqrt n$ with $n=44$. Since the measured velocities represent real motions of individual clouds, we adopt here the standard deviations as the errors of the mean velocity and distance.

We comment on possible systematic errors in the used parameters for calculation of kinematic distance. The Oort constant $A$ still includes uncertainty of $\sim 10$\%. The solar motion may include $\sim 1$ \kms uncertainty to cause systematic errors of several percents in LSR velocities. Also, interstellar turbulent motion may apply to the molecular clouds, yielding a few \kms uncertainties in kinematical velocities. These uncertainties may cause a systematic error of $\sim 20-30$\% in the derived distance.

\subsection{Comparison with optical extinction} 

Weaver (1949) showed that the color excess of stars toward Aquila Rift starts to increase at a distance modulus $m-M\sim 6$ mag to 8 mag, reaching a local maximum at $m-M\sim 8.5$ to 9. This indicates that the front edge of the dark cloud is at a distance of $r \sim 200$ pc, and attains maximum density at $\sim 500$ pc.  Straizys et al. (2003) showed that the front edge of Aquila rift is located at $r=225\pm 55$ pc, where the visual extinction $A_V$ starts to increase, and reaches a maximum at $r\sim 600$ pc. 

A more number of measurements are summarized in Dzib et al. (2010). Most of the measurements are toward the Serpens molecular cloud and/or core at $(l,b)= (31^\circ.4,~5^\circ.5)$. We reproduce their compilation in table 2,  where we added our new estimations using the data from the literature. The listed distances to Aquila Rift (Serpens cloud) are distributed from 200 pc to 650 pc. By averaging the listed values, we obtain a distance of $r=430 \pm 147$ pc.

Optical measurements yielded systematically smaller distances compared to the radio line kinematics and VLBI parallax measurements (Dzib et al. 2010). The latter  is consistent with the present CO line kinematical distance. The difference between optical and radio measurements would be due to selection effect of optical observations, in which stars are heavily absorbed so that farther stars are harder to be observed.  

\subsection{The molecular core of Aquila Rift}

From the spectrum we estimate the integrated CO intensity to be about $I_{\rm CO}\sim 20$ K \kms, which yields a column density of hydrogen molecules of $4\times 10^{21}{\rm H_2 cm^{-2}}$, or hydrogen atom column density of $8\times 10^{21}\hcm2$. This is about equal to the critical column density for R7 band, $N_7=9 \times 10^{21}\hcm2$, indicating that the R7-band emission is marginally absorbed by the Serpens clouds.

CO gas in the Aquila Rift is distributed from $b=0^\circ$ to 10$^\circ$ and from $l=20^\circ$ to $30^\circ$. For a mean distance to the ridge center of 0.642 kpc, the area is approximated by a rectangular triangle with one side length of $s\sim 100$ pc. Taking an average column density through the ridge of $N\sim 8 \times 10^{21}\hcm2$, we  estimate the cloud's mass to be $M\sim (s^2/2) m_{\rm H} N\sim 3\times 10^5 \Msun$. Hence, the Aquila rift is considered to be a giant molecular cloud of medium size and mass, enshrouded in an extended HI gas . 

If we assume a line of sight depth of the cloud to be $\sim 100$ pc, the volume density of gas is on the order of $\sim 26$ H cm$^{-3}$, lower than that of a typical giant molecular cloud. Taking a half velocity width $\sigma_v\sim 4$ \kms and radius $a \sim 50$ pc, the Virial mass is estimated to be on the order of $M_{\rm V} \sim \sigma_v^2 a/G \sim 2 \times 10^5\Msun$. Hence, the Aquila molecular cloud will be a gravitationally bound system, although the gravitational equilibrium assumption may not be a good approximation because of its complicated shape. 

In order to lift the cloud to the height of $h\sim r ~{\rm tan}~ 5^\circ \sim 50$ pc from the galactic plane, gravitational energy of $E_{\rm g}\sim 10^{50}$ ergs is required. The origin of Aquila rift in relation to its inflating-arch morphology lifted from the galactic plane would be an interesting subject for the future, particularly in relation to magnetic fields.

\begin{table}
\caption{Distances to the Aquila-Serpens rift$^\ddagger$.} 
\begin{tabular}{lll}  
\hline 
Authors & $r$ (pc) & Method \\
\hline
Weaver ('49)$^*$ & $450 \pm 100$ & $A_v$ opt  \\
Racine ('68) & 440 &   $A_v$ opt\\
Strom+ ('74) & 440 & $ A_v$ IR\\
Chavarria+ ('88) & $245 \pm 30$ & $A_v$ opt/IR\\
Zhang+ ('88) & $650 \pm 180$ & $A_v$ opt/IR \\
ibid & 600 & Kin. CO, NH$_3$\\ 
de Lara+ ('91) & $311 \pm 38$ & $A_v$ opt\\ 
Straizys+ ('96) & $259 \pm 37$ & $A_v$, front edge\\ 
ibid$^*$  & $500 \pm 50$ & $A_v$  \\
Dzib+ (2010)  & $415\pm 25$ & Maser parallax\\
Knude(2010, 11) & $203\pm 7$ & $A_v$ 2MASS, 1st peak \\
ibid$^*$ & $400 \pm  50$ & ibid, 2nd peak\\ 
This work $^*$ & $642\pm 174$ & Kin. $l=20$-$30^\circ, b=3$-$8^\circ$\\
Average $^*$  & $ 430\pm 147$\\
\hline
\end{tabular} 
\\
$\ddagger$ After Dzib et al. (2010) for Serpens cloud,\\
$*$ Our estimations.  \\
\end{table}

\section{The Distance of NPS} 

We now estimate a lower limit of the distance to NPS, assuming that the NPS is a part of a shell. The nearest allowed configuration of the shell with respect to the Aquila rift is illustrated in the lower panel of figure  10. In the nearest possible case shown by the dashed circle, the shell is in touch with the farthest edge of the Aquila rift. Here the shell center is assumed to be located in the direction of $l\sim 0^\circ$.  Let the distance of the tangential point of the shell  $r_{\rm t}$, the distance to the shell center $r_{\rm c}$, and the half thickness of the Aquila rift on the line of sight $a$. Then, the distance $r_{\rm t}$ and $r_{\rm c}$ are related to the distance of the Aquila Rift ridge $r$ as
\be
r_{\rm t}\simeq r+a+r'=(r+a)(1+{\rm sin}~l)
\ee
and 
\be
r_{\rm c}\simeq r_{\rm t}/{\rm cos}~l.
\ee

We take the determined distance of the molecular Aquila Rift, $r=0.642 \pm 0.174 ~{\rm kpc}$, and assume a line-of-sight width of 100 pc, or a radius $a \simeq 50$ pc, for the rift ridge at $ l \sim 27^\circ$. Then, we have a lower limit distance to the NPS as $r_{\rm t}\sim 1.01 \pm 0.25$ kpc, and to the shell center $r_{\rm c}\sim 1.13 \pm 0.29$ kpc. If we use the averaged value of measured distances to Aquila rift as listed in table 2, the lower limit distance to NPS would be farther than $0.70\pm 0.24$ kpc, and the shell center farther than $0.79\pm 0.27$ kpc. If the shell center is in the direction of $(l,b)\sim(330^\circ,20^\circ)$, as Loop I fitting suggests, the distance should be greater.

\begin{figure}  
\bc
\includegraphics[width=5cm]{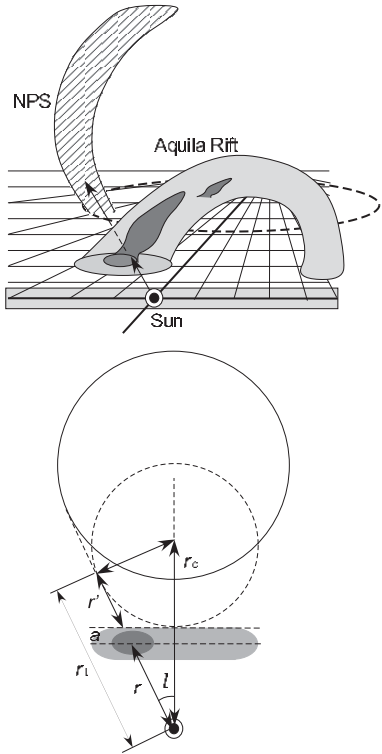}  
\ec
\caption{Schematic relationship between NPS and Aquila Rift. The dashed circle shows the nearest possible case for calculating a lower limit distance.}
\end{figure}

\begin{table} 
\caption{Lower-limit Distance of the North Polar Spur.}
\begin{tabular}{lll}  
\hline 
Method &Distance (kpc)\\
\hline
CO-line kin. dist. (This work) & $1.01 \pm 0.25$ kpc \\
Using the average in table 2 & $0.70 \pm 0.24$ kpc \\ 
Radio depolarization (Sun+ 2010)& $\sim 2-3$ kpc \\
Faraday rotation (This work) &$\sim 6 $ kpc  \\   
\hline 
\end{tabular}
\end{table}

\subsection{Comparison with Faraday distances of NPS} 

Sun et al. (2014) analyzed Faraday screening of linearly polarized radio continuum emission toward the low-latitude extension of NPS ridge. They concluded that the beam depolarization at 2.3 GHz indicates that the NPS ridge is located farther than 2 to 3 kpc.

They also showed that the B vectors at 4.8 GHz are aligned at position angles 
$\phi_1\simeq 40^\circ$ toward the most strongly polarized region at 
$(l,b)=(22^\circ.7, 4^\circ.0)$, whereas 2.3 GHz B vectors are aligned at 
$\phi_2=-40^\circ$. The RM value of extragalactic radio sources are 
$RM\sim +50$ to 
$-100$ rad m$^{-2}$ in the same direction of 
$l=23^\circ -25^\circ$, 
$b=3^\circ - 5^\circ$ (Taylor et al. 2009). Hence, we may choose a positive minimum rotation of the polarization angle to yield  
$\Delta \phi=+110^\circ$. This results in a rotation measure of
$ RM=\Delta \phi/(\lambda_2^2-\lambda_1^2)\sim +150 ~{\rm rad~m}^{-2}.$

If we take interstellar electron density of 
$n_{\rm e}\sim 10^{-2}~{\rm cm}^{-2}$ and magnetic field of 
$B\sim 3~\mu$G, the estimated RM corresponds to a line-of-sight depth of
$ r = |RM|/(0.82 n_{\rm e}B) \sim 6 ~{\rm kpc}.$  
The corresponding height is about 400 pc from the galactic plane, beyond which the Faraday rotation would be negligible. Hence, the here estimated value is a lower distance. In table 3 we summarize the estimated lower limits to the distance of NPS.

\subsection{Effect of the bulge}
Galactic bulge emission may be superposed on the used ROSAT maps, producing an absorption band along the galactic plane, particularly in the Galactic Center direction. Although the bulge emission in the analyzed region at 
$l\sim 20-30, ~b\sim 10-50 ^\circ$ has not been thoroughly investigated, it will be weaker than the NPS emission. Hence, the present result gives a lower limit to the X-ray emitting source of the NPS possibly superposed by weak bulge emission. However, this does not affect the present result, because the lower limit applies both to the distances of NPS and superposed emissions.

\section{Conclusion and Discussion}

\subsection{Summary}
We summarize the major results as follows:

(1) Soft X-rays from the NPS are absorbed by the galactic disk, specifically by dense clouds in the Aquila Rift, as evidenced by (A) the clear anticorrelation between R5 (0.89 keV) X-ray intensity and total H column density; and (B) the coincidence of R5-band intensity profile with the interstellar extinction profile, indicating that the soft X-rays obey the extinction law due to Aquila Rift gas.

(2) The kinematical distance to the absorbing gas in the Aquila Rift was measured to be $r=642\pm 174$ pc. Assuming a shell structure, the lower limit to the NPS ridge is estimated to be $1.02\pm 0.25$ kpc. 

(3) The derived distance shows that the major part of the NPS is located high above the galactic gas layer, and we conclude that the NPS is a galactic halo object. Since the gas density in the halo is too small to cause further absorption, the NPS distance is allowed to be much farther including the Galactic Center distance.

\subsection{Possible extinction-free NPS at low latitudes}

We may speculate about a possible intrinsic structure of NPS by correcting for the absorption by accepting the present distance estimation. We here use the ROSAT R7 data where the extinction is smallest among the available bands, so that overcorrection is less significant than in other bands. 
   
First, we smooth the R7 map by a Gaussian beam with $1^\circ$ FWHM in order to increase the signal-to-noise ratio to obtain figure 11a. Then, we divide the map by a map of ${\rm exp}(-\tau_i)={\rm exp}(-N/N_7)$ with $N_7=90.0 \times 10^{20} {\rm H cm}^{-2}$.

Figure 11 shows the thus obtained extinction-free R7 band map compared with the color-coded X-ray map from figure  1 . An X-ray spur is revealed to show up along at $b< 10^\circ$. Note that the region nearer to the galactic plane at $|b|<3^\circ$ may not be taken serious, where overcorrection due to the too large optical depth may exit. In figure 11(c) we compare the R7 map with a background-filtered 2300 MHz radio continuum map (Jonas et al. 1998). 

Both the radio and X-ray spurs emerge from the galactic plane at about the same angle $\sim 50^\circ$, and become sharper, narrower and brighter toward the plane. The X-ray spur is displaced to inside (westward) of the radio ridge by $\sim 5^\circ$, or $\sim 10-15$\% of the apparent curvature radius. The displacement is systematically observed over the entire NPS, and will be attributed to the difference of emitting mechanisms. We may also mention that the Fermi $\gamma$-ray bubble is located further inside the X-ray shell.

\subsection{Origin of NPS}
 
 If the NPS is a shock front of an old supernova remnant, the radio brightness-to-diameter relation applied to Loop I indicates a diameter of $\sim 100$ pc and a distance to NPS of the same order (Berkhuijsen 1971). An alternative model attributes the origin to a hypernova or multiple supernovae, or stellar wind from high-mass stars (Egger and Aschenbach 1995; Puspitarini et al.2014;  Wolleben 2007; Willingale et al. 2003). In this model the distance to the NPS is assumed to be $\sim 1$ kpc, which is consistent with the present lower-limit distance.  However, the model needs to explain the origin of an extraordinarily active star formation at altitudes as high as $\sim 300$ pc above the galactic disk.

The Galactic Center explosion model postulates bipolar hyper shells produced by a starburst 15 million years ago with explosive energy on the order of $10^{55-56}$ ergs in the Galactic Center (Sofue et al.  1977, 1984, 1994, 2000; Bland-Hawthorn et al. 2000). A shock front is supposed to reach a radius $\sim 10$ kpc in the polar regions and  $\sim 3$ kpc in the galactic plane. The lower limit distance, and the fact that the low-latitude narrow ridges in radio and X-rays emerge from the galactic plane are consistent with the Galactic Center explosion model.

\begin{figure}    
\bc
(a)\includegraphics[width=6cm]{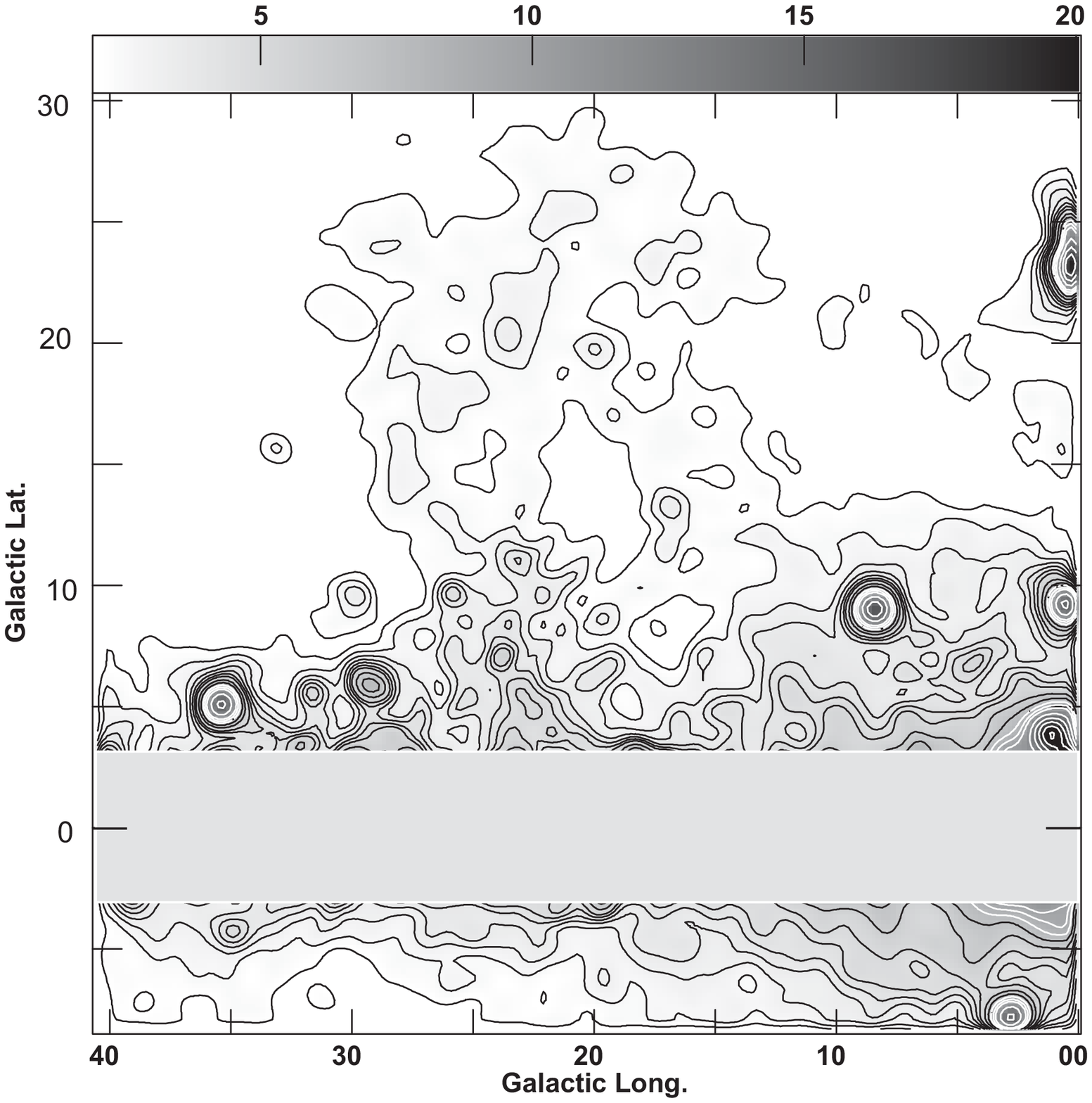}   \\
(b)\includegraphics[width=5.8cm]{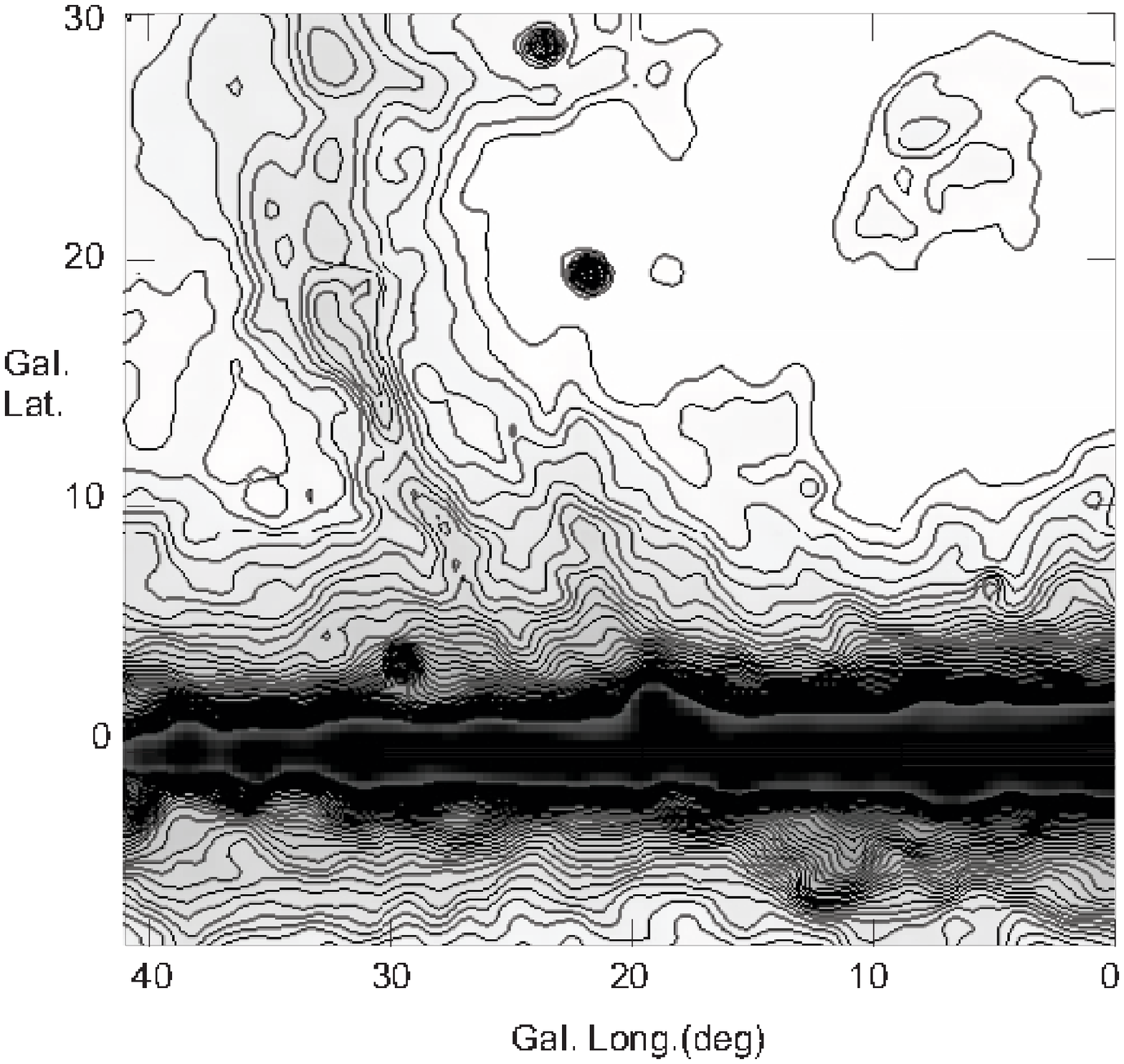}  
\ec  
\caption{ 
(a) Extinction-corrected R7 band X-ray map for the NPS region. Contours are from 70.4 by 17.6 count intervals in ROSAT count rate s$^{-1}$ arcmin$^{-1}$. (b) Background-filtered 2.3 GHz map from Rodhes survey (Jonas et al.1998) with contours every 0.05 Jy/beam. }
\end{figure}

\vskip 5mm \noindent{\it Acknowledgements}: 
 We thank the authors of the ROSAT all-sky survey, Leiden-Argentine-Bonn HI survey, and Columbia galactic CO line survey for providing us with the archival data.

\end{document}